\documentclass[twocolumn,a4paper]{article}
\usepackage{epsfig}
\usepackage[reqno]{amstex}
\normalsize
\topmargin \paperheight \advance\topmargin by -\textheight
\divide\topmargin by 2 \advance\topmargin by -1in
\oddsidemargin \paperwidth \advance\oddsidemargin by -\textwidth
\divide\oddsidemargin by 2 \advance\oddsidemargin by -1in
\evensidemargin \oddsidemargin

\makeatletter
\renewcommand{\@makecaption}[2]{%
  \vskip\abovecaptionskip
  {\small
     \sbox\@tempboxa{#1: #2}%
     \ifdim \wd\@tempboxa >\hsize
       #1: #2\par
     \else
       \hbox to\hsize{\hfil\box\@tempboxa\hfil}%
     \fi
  }
  \vskip\belowcaptionskip%
}
\makeatother

\makeatletter
\renewenvironment{thebibliography}[1]%
{%
   \list{\@biblabel{\arabic{enumiv}}}%
   {\settowidth\labelwidth{\@biblabel{#1}}%
   \leftmargin\labelwidth
   \advance\leftmargin\labelsep
\@openbib@code
      \advance\leftmargin\bibindent
      \itemindent -\bibindent
      \listparindent \itemindent
      \parsep \z@
   \usecounter{enumiv}%
   \let\p@enumiv\@empty
   }%
      \renewcommand\newblock{\par}
   \sloppy\clubpenalty4000\widowpenalty4000%
   \sfcode`\.=\@m%
}%
{%
   \def\@noitemerr
   {\@latex@warning{Empty `thebibliography' environment}}%
   \endlist%
}
\renewcommand\newblock{\hskip .11em \@plus.33em \@minus.07em}%
\makeatother
\begin{document}
\begin{flushleft}\bfseries\large
  Comment on ``Monte Carlo Algorithms with Absorbing Markov Chains:
  Fast Local Algorithms for Slow Dynamics`` 
\end{flushleft}\medskip

In \cite{novo} the problem of non-critical slowing of Monte-Carlo
simulations is addressed. This phenomenon may occur for instance in
low-temper\-ature simulations of the Metropolis or Glauber dynamics if
the system gets stuck in a local minimum of the energy. To get around
this problem, a different algorithm for simulation of the same
dynamics has been presented in \cite{boka}. This algorithm, commonly
called the $n$-fold way, directly simulates the times the system is at
rest. It is most efficient if the attraction basin of the energy's
local minima are small, that is, if the probability to leave a basin
after one step and afterwards to stay outside for a long period of time
is high.

It may however happen that the attraction basins are of moderate size,
for example if some energetically degenerate configurations are in the
same basin. Thus it is reasonable to construct a generalisation of the
$n$-fold-way algorithm, which simulates the sojourn times in the
basins and the exiting configurations without altering the given
dynamics.

One such algorithm is proposed in \cite{novo}. This algorithm,
however, contains one element which changes the dynamics drastically
and -- even worse -- it does not produce the correct stationary
distribution. Let us illustrate this by the following simple
example. Consider a random walk $\xi$ in $\{0,1,2,\dots,100\}$ with
reflecting boundaries at $0$ and $100$ and probability $\frac{1}{2}$
to move left or right at each time step inside. The stationary
distribution of $\xi$ is $\mathbb{P}(\xi=n)=\frac{1}{100}$ for $1\le
n\le 99$ and $\mathbb{P}(\xi=n)=\frac{1}{200}$ for $n=0,100$. To every
state $n$ we associate a basin $\mathfrak{B}_n:=\{n\}$ if $n\le 40$ or
$n\ge 60$ and $\mathfrak{B}_n:=\{n,n+1\}$ if $40<n<60$. For this
setting we have performed a simulation according to the recipe given
in \cite{novo}. The algorithm produces a single limiting distribution
depicted in Fig.~1, which is clearly distinct from the correct one. It
effectively introduces a spurious drift depending on the basins
chosen. This will also change dynamical properties.

We finally give the relevant probabilities from which a correct
algorithm may readily be constructed in the spirit of \cite{novo}.
Consider a Markov chain $\Sigma$ in the configuration space with given
one time step transition probability
$\mathbb{P}(\Sigma(t+1)=\sigma'\,\vert\,\Sigma(t)=\sigma)
=p_1(\sigma'\,\vert\,\sigma)$. To every configuration $\sigma$
associate a set
$\mathfrak{B}_\sigma=:\{\tilde\sigma_1,\dots,\tilde\sigma_{b_\sigma}\}$ of
$b_\sigma\ge 1$ basin configurations containing at least
$\sigma=:\tilde\sigma_1$ itself. With the help of the $b_\sigma\times
b_\sigma$-matrix $B_{\sigma}:=
(p_1(\tilde\sigma_k\,\vert\,\tilde\sigma_l))_{k,l=1,\dots,b_\sigma}$
the conditional distribution of the first exit time
\begin{equation}
  T_t := \min\{ t+\tau:\Sigma(t+\tau)\not\in\mathfrak{B}_{\Sigma(t)};
  \tau\ge 1\}
\end{equation}
from the basin can be written as
\begin{equation}
   \mathbb{P}(T_t=t+\tau\,\vert\,\Sigma(t)=\sigma)
   = \beta^\dagger(1-B_\sigma^{\phantom{\tau}})
     B_\sigma^{\tau-1}\delta,
\end{equation}
where we have used the $b_\sigma$-vectors
$\beta^\dagger:=(1,1,\dots,1)$, $\delta^\dagger:=(1,0,0,\dots,0)$ and
the dagger denotes transposition. The distribution of the exit times
is in accordance with that stated in \cite{novo}. Now form the set
$\mathfrak{E}_\sigma:=\{\hat\sigma_1,\dots,\hat\sigma_{e_\sigma}\}$ of all
$e_\sigma$ exit configurations outside $\mathfrak{B}_\sigma$ that can
be reached from $\mathfrak{B}_\sigma$ in one time step. The
conditional probability to be in one of these configurations at given
first exit time is
\begin{multline}\label{true}
   \mathbb{P}(\Sigma(T_t)=\hat\sigma_n
              \,\vert\,T_t=t+\tau;\Sigma(t)=\sigma) \\
   = \frac{
           \varepsilon^\dagger_n
           E_\sigma^{\phantom{\tau}} B_\sigma^{\tau-1} \delta
         }{
           \beta^\dagger(1-B_\sigma)
           B_\sigma^{\tau-1}\delta
          },
\end{multline}
where we have introduced the $e_\sigma\times b_\sigma$-matrix
$E_\sigma:=(p_1(\hat\sigma_n\,\vert\,\tilde\sigma_k)
)_{n=1,\dots,e_\sigma;k=1,\dots,b_\sigma}$ and the $e_\sigma$-vector
$(\varepsilon_n)_m:=0$ for $m\not=n$, $(\varepsilon_n)_n:=1$. The
algorithm proposed in \cite{novo} would correspond to substitute
the right-hand side of (\ref{true}) by
\begin{equation}\label{false}
   \frac{
         \sum_{\tau'=0}^{\tau-1}
         \varepsilon^\dagger_n
         E_\sigma^{\phantom{\tau}} B_\sigma^{\tau'} \delta
       }{
         \sum_{\tau'=0}^{\tau-1}
         \beta^\dagger(1-B_\sigma^{\phantom{\tau}})
         B_\sigma^{\tau'}\delta
       }.
\end{equation}
\begin{figure}[t]
\epsfig{file=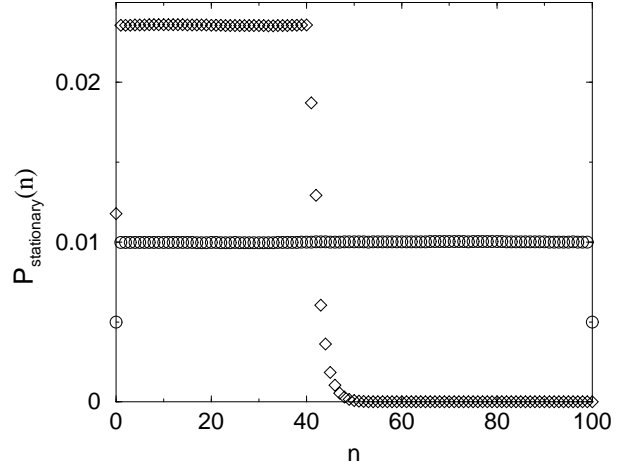, height=70mm, width=84mm}
\caption[0]{%
Distribution of the random walk described in the text using
Eq. (\ref{true}) ({\large $\circ$}) or Eq. (\ref{false}) ({\large
$\diamond$}) averaged over $10^7$ time steps starting from a uniform
initial distribution.%
}
\end{figure}
\bigskip

\noindent
{\large K.~Broderix and R.~Kree}

{\small
Georg-August-Universit\"at G\"ottingen

Institut f\"ur Theoretische Physik

Bunsenstr. 9, D-37073 G\"ottingen

Germany

\makeatletter
Email: broderix$\,$@$\,$theorie.physik.uni-goettingen.de
\makeatother
}
\bigskip

\noindent
Version of June 28, 1995

\medskip


\end{document}